\documentclass[superscriptaddress,showpacs,aps,prd]{revtex4-1}
\usepackage{epsfig}
\usepackage{amsmath,amssymb}
\usepackage{color}

\begin{document}
\title{Conformastatic disk-haloes in Einstein-Maxwell gravity}

\author{Antonio C. Guti\'errez-Pi\~{n}eres}
\email[e-mail:]{acgutierrez@correo.nucleares.unam.mx}
\affiliation{Facultad de Ciencias B\'asicas,
Universidad Tecnol\'ogica de Bol\'ivar, Cartagena, Colombia}
\affiliation{Instituto de Ciencias Nucleares, Universidad Nacional Aut\'onoma 
de M\'exico, AP 70543,  M\'exico, DF 04510, M\'exico}

\author{Guillermo A. Gonz\'alez} 
\email[e-mail:]{guillego@uis.edu.co}
\affiliation{Escuela de F\'{\i}sica, Universidad Industrial de Santander, A. A. 678,
Bucaramanga, Colombia}

\author{Hernando Quevedo}
\email[e-mail:]{quevedo@nucleares.unam.mx}
\affiliation{Instituto de Ciencias Nucleares, Universidad Nacional Aut\'onoma 
de M\'exico, AP 70543,  M\'exico, DF 04510, M\'exico}
\affiliation{Instituto de Cosmologia, Relatividade e Astrofisica ICRA - CBPF
Rua Dr. Xavier Sigaud, 150, CEP 22290-180, Rio de Janeiro, Brazil}
\affiliation{Dipartimento di Fisica and Icra, Universit\`a di Roma "La Sapienza", I-00185
Roma, Italy}

\begin{abstract} 
We present a relativistic model describing a thin disk surrounded by a halo in presence of
an electromagnetic field. The model is obtained by solving the Einstein-Maxwell equations on a particular
conformastatic spacetime background  and by using the distributional approach for the energy-momentum tensor.
A class of solutions is obtained in which the gravitational and electromagnetic 
potentials are completely determined by a harmonic function only. A particular solution is given 
that is asymptotically flat and singularity-free, and satisfies all the energy conditions.  
\end{abstract}


\maketitle

\section{\label{sec:intro}Introduction}

A large number of galaxies and other  astrophysical systems have extended mass distributions
surrounded by a material halo.  For practical reasons one can assume that  many of these
systems preserve axial symmetry and, therefore, they can be modeled  in terms of
relativistic thin disks with exterior halos,  in particular when the gravity field is strong  
enough. Disks may also be used to model accretion disks, galaxies in thermodynamic
equilibrium and the superposition of a black hole and a galaxy. Disk sources for
stationary axially symmetric spacetimes with magnetic fields are also of astrophysical
importance  mainly in the study of neutron stars, white dwarfs and galaxy formation. 
To describe the gravitational and electromagnetic fields of such configurations, we will 
use general relativity and Maxwell's theory. Consequently, we are interested in deriving and
analyzing exact solutions of the Einstein-Maxwell equations. 
On
the other hand, the study of axially symmetric solutions of the Einstein and
Einstein-Maxwell field equations corresponding to shells and disk-like configurations of
matter, apart from its astrophysical relevance, has a clear purely mathematical interest.

Exact solutions that have relativistic static thin disks as their sources were first
studied by Bonnor and Sackfield \cite{BS} and Morgan and Morgan \cite{MM1,MM2}.
Subsequently, several classes of exact solutions corresponding to static
\cite{VOO,LP1,CHGS,LO,LEM,BLK,BLP,GL1,GE,GG-PV} and stationary
\cite{LP2,BL,PL,GL2,GG-Pcqg} thin disks have been obtained by different authors. The
superposition of a static or stationary thin disk with a black hole has been
considered in \cite{LL1,LL2,LL3,SZ1,SEM1,SZ2,SEM2,SEM3,KHZ}. Thin disks around static
black holes in a magnetic field have been studied in \cite{G-PG-RG}. Relativistic disks
embedded in an expanding Friedman-Lema\^itre-Roberton-Walker universe have been studied in \cite{FIL}, and perfect fluid disks
with halos in \cite{VL1}. Furthermore, the stability of thin disks models has been investigated
using a first order perturbation of the energy-momentum tensor in \cite{UL1}. On the other
hand, thin disks have been discussed as sources for Kerr-Newman fields \cite{LBZ,GG1},
magnetostatic axisymmetric fields \cite{LET1, G-PGijtp}, and conformastatic and
conformastationary metrics \cite{VL2,KBL,GG-PO}. Also, models of electrovacuum static
counterrotating dust disks were presented in \cite{GG2}, charged perfect fluid disks were
studied in \cite{VL3}, and charged perfect fluid disks as sources of static and
Taub-NUT-type spacetimes in \cite{GG3,GG4}. Also, monopole and dipole layers in curved
spacetimes were analyzed in \cite{GBG-P1}, and electromagnetic sources distributed on shells in a
Schwarzschild background in \cite{GBG-P2}.

Now, the thin disks with magnetic fields presented in \cite{LBZ,GG1,LET1} were
obtained by means of the well-known ``displace, cut and reflect'' method that 
introduces a discontinuity in the first-order derivative of an otherwise smooth solution. The
result is a solution with a singularity of the delta-function type in the entire $z = 0$
hypersurface, and so it can be interpreted as an infinite thin disk. On the other hand,
solutions that can be interpreted as thin disks of finite extension can be obtained if an appropriate
coordinate system is introduced. A coordinate system that adapts naturally to a
finite source and presents the required discontinuous behavior is given by the oblate
spheroidal coordinates. Some examples of finite thin disks obtained from vacuum solutions expressed
in these coordinates can be found in references \cite{BS,MM1,VOO,LO}, and from
electrovacuum solutions in reference \cite{GG-PO}.

In a previous work \cite{GG-PO}, we presented an infinite family of conformastatic axially
symmetric charged dust disks of finite extension with well-behaved surface energy and
charge densities. These disks have a charge density that is equal, up to a sign, to their
energy density, and so they are examples of the commonly named ``electrically counterpoised
dust'' equilibrium configuration. The energy density of the disks is everywhere positive
and well-behaved, vanishing at the edge. Furthermore, since the energy density of the disks
is everywhere positive and the disks are made of dust, all the models are in a complete
agreement with all the energy conditions, a fact of particular relevance in the study of
relativistic thin disks models. In the present paper, we extend these studies to
obtain a model corresponding to  a system that is composed of a thin disk and an exterior
halo. The main purpose of this  work is, then, to extend the previous electric field to include an
electromagnetic  field, and the previous ``isolated'' thin disk to include a thin disk-halo system.

In this work, we present a relativistic model describing a thin disk
surrounded by a halo in presence of an electromagnetic field in a conformastatic spacetime \cite{SYN,KRAMER,GV}. Note that we take the definition in \cite{KRAMER} as standard,
following the original terminology by Synge \cite{SYN}:
conformastationary are those stationary spacetimes
with a conformally flat space of orbits
and the conformastatic comprise the static subset. The model is obtained by
solving the Einstein-Maxwell equations 
 and by using the distributional
approach under the assumption that the energy-momentum tensor can be expressed as the sum of two
distributional contributions, one due to the electromagnetic part and the other one due
to a ``material'' part. In this way, explicit expressions for the energy,
pressure,
electric current and electromagnetic field are obtained for the disk region and for the
halo region. In order to obtain the solutions, an auxiliary function is introduced that determines
the functional dependence of the metric  and the
electromagnetic potential. It is also assumed that the auxiliary function depends 
explicitly on an additional function which is taken as a solution of the Laplace
equation. A simple thin disk-halo model is obtained from the Kuzmin solutions of the
Laplace equation. The energy-momentum tensor of the system agrees with all the  energy
conditions.

The plan of our paper is as follows. First, in Section \ref{sec:einm},  the conformastatic
line element is considered. The procedure to obtain electromagnetostatic, axially symmetric, relativistic
thin disks surrounded by a material halo is also summarized in this section. Section
\ref{sec:dhk} introduces a functional relationship dependence
between the metric and electromagnetic potentials and an auxiliary function in order to
obtain a family of solutions of  the  Einstein-Maxwell equations  in terms of  convenient
solutions of the Laplace equations, modeling relativistic thin disk-halo systems. Next, the
eigenvalue problem for disk with halos is studied and  a particular model of  a disk with
halo is obtained from the Kuzmin solutions of the Laplace equation. In Section \ref{sec:beh},
the  quadratic Riemann invariants and the electromagnetic invariants are studied, and the
behavior of the Kuzmin disk with halo is analyzed. Finally,  Section \ref{sec:conc}, is
devoted to a discussion of the results.

\section{The Einstein-Maxwell equations and the thin-disk-halo system}\label{sec:einm}
In order to formulate the Einstein-Maxwell equations for conformastatic axially symmetric
spacetimes corresponding to an electromagnetized system, constituted by a thin disk and a
halo surrounding the exterior of the disk, we  first  introduce coordinates
$x^a=(t,\varphi,r,z)$ in which the metric tensor  and the electromagnetic potential  only
depend on $r$ and $z$. We assume that these coordinates are quasicylindrical in the sense
that the coordinate $r$ vanishes on the axis of  symmetry  and, for fixed $z$, increases
monotonically  to infinity, while the coordinate $z$, for $r$ fixed,  increases
monotonically  in the  interval $(-\infty, \infty)$. The azimuthal angle $\varphi$ ranges
in the interval $[0,2\pi)$, as usual \cite{MM1, MM2}. We assume that there exists an
infinitesimally thin disk, located at the hypersurface $z=0$, so that the  components of
the metric tensor $g_{ab}$ and the components of the electromagnetic potential $A_a$ are
symmetrical functions of $z$ and their first derivatives  have a finite discontinuity at
$z=0$, accordingly,
\begin{eqnarray}
 g_{ab}(r,z)=g_{ab}(r,-z), \qquad A_{a}(r,z)=A_{a}(r,-z) 
\end{eqnarray}
in such a way that, for $z\neq 0$,
\begin{eqnarray}
 g_{ab,z}(r,z)=-g_{ab,z}(r,-z),\qquad A_{a,z}(r,z)= -A_{a,z}(r,-z).
\end{eqnarray}
The metric  tensor and the  electromagnetic  potential are continuous at $z=0$,
\begin{subequations}
\begin{eqnarray}
\left[g_{ab}\right] &=& g_{ab}|_{_{z = 0^+}} -  g_{ab}|_{_{z = 0^-}} = 0,\\
&\nonumber\\
\left[A_a\right] &=& A_{a}|_{_{z = 0^+}} -  A_{a}|_{_{z = 0^-}} = 0,
\end{eqnarray}
\end{subequations}
whereas the  discontinuity in the derivatives of the metric tensor and the  electromagnetic  potential can be  written,
respectively, as
\begin{subequations}
\begin{eqnarray}
\gamma_{ab} &=&  [{g_{ab,z}}],\\
&\nonumber\\
\zeta_{a} &=& [{A_{a,z}}],
\end{eqnarray}
\end{subequations}
where the reflection symmetry with respect to $z=0$ has been used. 
Then, by using the  distributional approach \cite{PH,LICH,TAUB} or the junction
conditions on the extrinsic  curvature of thin shell \cite{IS1,IS2,POI}, we  can write
the metric and  the electromagnetic potential as
\begin{subequations}
\begin{eqnarray}
g_{ab} &=& g^+_{ab} \theta (z) + g^-_{ab} \{ 1 - \theta (z) \},\\
&\nonumber\\
A_{a} &=& A^+_{a} \theta (z) + A^-_{a} \{ 1 - \theta (z) \},
\end{eqnarray}
\end{subequations}
and thus the Ricci tensor reads
\begin{equation}
R_{ab} = R^+_{ab} \theta(z) + R^-_{ab} \{ 1 - \theta (z) \} + H_{ab} \delta(z),
\label{eq:ricdis}
\end{equation}
where $\theta(z)$ and $\delta (z)$ are, respectively, the Heaveside and Dirac
distributions with support on $z = 0$. Here $g^\pm_{ab}$ and $R^\pm_{ab}$ are
the metric tensors and the Ricci tensors of the $z \geq 0$ and $z \leq 0$
regions, respectively, and
\begin{eqnarray}
H_{ab} = \frac{1}{2} \{ \gamma^z_a \delta^z_b  + \gamma^z_b \delta^z_a
-\gamma^c_c \delta^z_a \delta^z_b - g^{zz} \gamma_{ab} \},
\end{eqnarray}
where all the quantities are evaluated at $z = 0^+$. In agreement with Eq.(\ref{eq:ricdis}),
the energy-momentum tensor and the electric current density can be expressed as 
\begin{subequations}
\begin{eqnarray}
T_{ab} &=& T^+_{ab} \theta(z) + T^-_{ab} \{ 1 - \theta(z) \} + Q_{ab} \delta(z),
\label{eq:emtot}\\
& \nonumber\\
J_a &=& J^+_{a} \theta(z) + J^-_{a} \{ 1 - \theta(z) \} + I_{a} \delta(z),
\label{eq:eccomp}
\end{eqnarray}
\end{subequations}
where $T^\pm_{ab}$ and $J^\pm_{ab}$ are the energy-momentum tensors and electric current
density of the $z \geq 0$ and $z \leq 0$ regions, respectively. Moreover, $Q_{ab}$ and $I_{a}$
represent the part of the energy-momentum tensor and the electric current density
corresponding to the disk-like source. 

To describe the  physical properties of an electromagnetized system constituted by a thin
disk  surrounded by an exterior halo, $T^\pm_{ab}$ in (\ref{eq:emtot}) can be written as
\begin{eqnarray}
 T^\pm_{ab} = E^\pm_{ab} + M^\pm_{ab},\label{eq:emtotcomp}
\end{eqnarray}
where $E^\pm_{ab}$ is the electomagnetic energy-momentum tensor
\begin{eqnarray}
E_{ab} = F_{ac}F_b^{ \ c} - 
\frac{1}{4}
g_{ab}F_{cd}F^{cd}, \label{eq:tab}
\end{eqnarray}
with $F_{ab} =  A_{b,a} -  A_{a,b}$ and $M^\pm_{ab}$  is  an unknown ``material''
energy-momentum tensor (MEMT) to be obtained. Accordingly, the Einstein-Maxwell
equations, in geometrized units such that $c = 8\pi G = \mu _{0} = \epsilon _{0} =  1$,
are equivalent to the system of equations
\begin{eqnarray}
G_{ab}^{\pm} =R^\pm_{ab} - \frac{1}{2} g_{ab} R^\pm &=& E^\pm_{ab} + M^\pm_{ab},
\label{eq:einspm}
\\
H_{ab} - \frac{1}{2} g_{ab} H &=& Q_{ab}, \label{eq:einsdis}\\
\hat F^{ab}_{ \pm  \ \ , b}  &  =  & \hat J^{a}_\pm ,\label{eq:maxext}\\
\  [\hat F^{ab}]n_{_{b}} & = & \hat I^{a},\label{eq:emcasj}
\end{eqnarray}
where $H = g^{ab} H_{ab}$ and $\hat a = \sqrt{-g}a$, being $g$ the determinant of the
metric tensor. Here, ``$[\;\;]$" in  the expressions $[\hat F^{ab}]$ denotes the jump of
$\hat F^{ab}$
across of the surface $z=0$ and $n_{_{b}}$ denotes an unitary vector in the  direction
normal  to it.



Now, in order to obtain explicit forms for the  Einstein-Maxwell equations
corresponding to the electromagnetized disk-halo system, we take the metric tensor as
given by the conformastatic line element \cite{SYN, KRAMER}
\begin{eqnarray}
 \mathrm ds^2 = - \mathrm e^{2 \phi} \mathrm dt^2 \ + \mathrm e^{2 \psi}
 [r^2{\mathrm d}\varphi^2  + \mathrm dr^2 + \mathrm
 dz^2], \label{eq:met0} 
\end{eqnarray}
where the metric functions $\phi$ and $\psi$ depend only on $r$ and $z$, and the electromagnetic potential is
\begin{equation}
A_{\alpha} = (A_0, A, 0, 0).
\end{equation}
We also assume that the electric potential $A_0$ and the magnetic potential $A$
are independent of $t$. For later use, the corresponding field equations and the components of the energy-momentum
tensor are given explicitly in the Appendix. 

Furthermore, from Eq.(\ref{eq:maxext}) we have for the current density in these regions,
\begin{subequations}\begin{eqnarray}
&&\hat{J}_{\pm}^{0} = -re^{\psi -\phi}\{\nabla^2A_{0} -\nabla(\phi -\psi)\cdotp \nabla A_0\},\\
&& \hat{J}_{\pm}^{1} = -r^{-1}e^{ \phi - \psi}\{\nabla^2A + \nabla(\phi - \psi)\cdotp\nabla A 
 - \frac{2}{r}A_{,r}\},
\end{eqnarray}\label{eq:current2}\end{subequations}
where, as we know, $\hat{J}_{\pm}^{a}= {J}_{\pm}^{a}\sqrt{-g}$. The ``true'' surface
energy-momentum tensor of the disk, $S_{ab}$, can be obtained through the
relation
\begin{equation}
S_{ab} \ = \ \int Q_{ab} \ \delta(z) \ ds_n \ = \ \sqrt{g_{zz}} \ Q_{ab} ,
\end{equation}
where $ds_n = \sqrt{g_{zz}} \ dz$ is the ``physical measure'' of length in the
direction normal to the $z = 0$ plane. Accordingly, for the  metric  (\ref{eq:met0}), the
nonzero components of $S_{ab}$ are given by
\begin{subequations}\begin{eqnarray}
S^0_{\;0} &=& 4 e^{-\psi}\psi_{,z},\\
S^1_{\;1} &=& 2 e^{-\psi}(\phi + \psi)_{,z},\\
S^2_{\;2} &=& 2 e^{-\psi}(\phi + \psi)_{,z},
\end{eqnarray}\label{eq:surfacemt}\end{subequations}
where all the quantities are evaluated at $z=0^+$. The ``true'' current density on the
surface of the disk, ${\cal J}_a$, can be obtained through the  relation
\begin{eqnarray}
  {\cal J}^{a} = \int{I^{a}\delta(z)\sqrt{g_{zz}}dz},\label{eq:currentgen}
\end{eqnarray}
or explicitly 
\begin{subequations}
\begin{eqnarray}
 {\cal J}^0= e^{-(\psi + 2\phi)}[A_{0,z}], \\
  {\cal J}^1= - r^{-2}e^{-3\psi}[A_{,z}],
\end{eqnarray}
\end{subequations}
where all the quantities are evaluated on the surface of the disk, and
$[A_{a,z}]$ denotes the jump of the derivative of $A_a$ across  the surface $z=0$. Now, in
order to  analyze the physical characteristics of the  system it  is convenient to express
the energy-momentum tensor $T_{ab}$ and the electric current density in terms of an
orthonormal tetrad.  We  will use the tetrad of the ``locally static  observers'' (LSO)
\cite{KBL}, i.e., observers at rest with respect to infinity, which is given by 
\begin{eqnarray}
e_{(b)}^{\;\;a}=\{V^a,W^a,X^a,Y^a\},
\end{eqnarray}
 where
\begin{subequations}\begin{eqnarray}
V^{a} &=& e_{(0)}^{\;\;a}= e^{-\phi} \delta_{0}^{a}, \\ 
W^{a} &=& e_{(1)}^{\;\;a}= r^{-1}e^{-\psi} \delta_{1}^{a}, \\ 
X^{a} &=& e_{(2)}^{\;\;a}= e^{-\psi} \delta_{2}^{a},\\ 
Y^{a} &=& e_{(3)}^{\;\;a}= e^{-\psi} \delta_{3}^{a}.
\end{eqnarray}\end{subequations}
In terms of this tetrad $M_{ab}^{\pm}$ and  $\hat{J}^{\pm}_{a}$  can be expressed as  
\begin{subequations}
\begin{eqnarray}
M_{\pm}^{ab}&=&M^{\pm}_{(0)(0)}V^aV^b + M^{\pm}_{(1)(1)}W^aW^b +M^{\pm}_{(2)(2)} X^aX^b
+ M^{\pm}_{(3)(3)}Y^aY^b \\&-&M^{\pm}_{(0)(1)}\left\{V^aW^b +W^aV^b\right\} +
M^{\pm}_{(2)(3)}\left\{X^aY^b + Y^aX^b\right\}\nonumber,\\
\hat{J}_{\pm}^{a} &=& - \hat{J}^{\pm}_{(0)}V^a + \hat{J}^{\pm}_{(1)}W^a.
\end{eqnarray}
\end{subequations}
The explicit expressions for these quantities are given in the Appendix.

In the  same way, by using the  LSO tetrad, the surface energy-momentum tensor 
and the surface current density of the disk  as well as the electric current on the disk can be
written in the canonical form as
\begin{subequations}\begin{eqnarray}
 S^{ab} &=& S_{(0)(0)}V^{a}V^{b} + S_{(1)(1)}W^{a}W^{b} +
S_{(2)(2)}X^{a}X^{b},\\
{\cal {\cal J}}^{a} &=& - {\cal J}_{(0)}V^{a} + {\cal J}_{(1)}W^{a},
\end{eqnarray}\end{subequations}
with
\begin{subequations}\begin{eqnarray}
S_{(0)(0)} &=& -4e^{-\psi}\psi_{,z},\\
S_{(1)(1)} &=& 2e^{-\psi}(\phi + \psi)_{,z} = S_{(2)(2)},\\
{\cal J}_{(0)} &=& - e^{-(\psi + \phi)}[A_{0,z}],\\
{\cal J}_{(1)} &=& -ṛ^{-1}e^{-2\psi}[A,z],
\end{eqnarray}\label{eq:surcorr}\end{subequations}
where we have  used (\ref{eq:surfacemt}) and (\ref{eq:currentgen}) and  again all the
quantities are evaluated at
$z=0^+$.
\section{Thin disk with an electromagnetized material halo}
\label{sec:dhk}

In the precedent section, we discussed a generalized formalism in which  conformastatic
axially symmetric  solutions of the Einstein-Maxwell  can be  interpreted in terms of a
thin disk placed at the surface $z=0$ surrounded by a distribution of   electrically charged matter
located in the $z \geq 0$ and $z \leq 0$ regions, whose physical properties can be studied
by analyzing the behavior of $S_{(a)(b)}$, ${\cal J}_{(a)}$, $M_{(a)(b)}^{\pm}$ and ${J}_{(a)}^{\pm}$. 
To this end, it is necessary to ``choose'' a convenient explicit form for
the metric. 
It turns out that the assumption $\psi = -\phi$ leads to a  
considerable simplification of the problem. Indeed, the equations
(\ref{eq:MJtet}) reduce to 
\begin{widetext}
\begin{subequations}\begin{eqnarray}
{M_{(0)(0)}^{\pm}} &=&\frac{1}{4}f^{-1}
\{4f\nabla^2f -5\nabla f \cdot \nabla f -2f \nabla A_0 \cdot \nabla A_0
-2r^{-2}f^3 \nabla A \cdot \nabla A\},\\
{M_{(0)(1)}^{\pm}} &=& -r^{-1}f \nabla A_0 \cdot \nabla
A,\\
{M_{(1)(1)}^{\pm}} &=&\frac{1}{4}f^{-1}
\{\nabla{f}\cdot\nabla{f} - 2f\nabla{A_0}\cdot\nabla{A_0}
 -2r^{-2}f^3 \nabla{A}\cdot\nabla{A}\},\\
{M_{(2)(2)}^{\pm}} &=&
\frac{1}{4}f^{-1}\{-(f_{,r}^2 - f_{,z}^2) + 2f(A_{0,r}^2 - A_{0,z}^2) -
2r^{-2}f^3(A_{,r}^2 - A_{,z}^2)\},\\
{M_{(3)(3)}^{\pm}} &=&-\overset{\mbox {\tiny
M}}{M_{(2)(2)}^{\pm}},\\
{M_{(2)(3)}^{\pm}} &=&
-\frac{1}{2}f^{-1}f_{,r}f_{,z} + A_{0,r}A_{0,z} - r^{-2}f^2A_{,r}A_{,z},
\label{eq:nondiagm23}\\
 \hat J_{(0)}^{\pm} &=& rf^{1/2}
\nabla\cdot(f^{-1}\nabla{A_0})\label{eq:current0},\\
 \hat J_{(1)}^{\pm} &=& -r^2f^{-1/2}
\nabla\cdot(r^{-2}f\nabla{A})\label{eq:current1}.
\end{eqnarray}\label{eq:emth2m}\end{subequations}\end{widetext}

In the  same way, for the nonzero components of the energy-momentum and the 
current density on the surface of the  disk (\ref{eq:surcorr})  we have, respectively,
\begin{eqnarray}
 S_{(0)(0)}=4(f^{1/2})_{,z},\label{eq:stet}
\end{eqnarray}
 and
\begin{subequations}\begin{eqnarray}
  {\cal J}_{(0)}&=&-[A_{0,z}],\\
  {\cal J}_{(1)}&=&-r^{-1}f[A_{,z}],
\end{eqnarray}\label{eq:currtet}\end{subequations}
where ``$[\;\;]$'' denotes the  jump across of the disk, $f\equiv e^{2\phi}$ and all the
quantities are evaluated on the surface of the  disk. We will suppose that there is no
electric current in the  halo, i. e., we assume that $\hat J_{(\alpha)}^{\pm} \equiv 0$.
Then, if $\hat{\mathbf e}_{\varphi}$ is a unit vector in the azimuthal direction and
$\lambda$ is  any reasonable function independent of the azimuth, one has the identity
\begin{eqnarray}
 \nabla\cdot(r^{-1}\hat{\mathbf
e}_{\varphi}\times\nabla\lambda) = 0\label{eq:identity}.
\end{eqnarray}
Equation (\ref{eq:current1}) may be regarded as the integrability condition for
the  existence of the  function $\lambda$ defined by
\begin{eqnarray}
 r^{-2}f\nabla{A}=r^{-1}\hat{\mathbf
e}_{\varphi}\times\nabla\lambda,\label{eq:gradalamb}
\end{eqnarray}
or, equivalently
\begin{eqnarray}
 -f^{-1}\nabla\lambda= r^{-1}\hat{\mathbf
e}_{\varphi}\times\nabla{A}.
\end{eqnarray}
Hence, the identity (\ref{eq:identity}) implies the  equation
\begin{eqnarray}
 \nabla\cdot(f^{-1}\nabla\lambda)=0\label{eq:auxiliar}
\end{eqnarray}
for the new ``potential'' $\lambda(r,z)$. In order to obtain an explicit form of  the
metric and  electromagnetic  potential, we suppose that $f, A_0$ and $A$ depend explicitly on 
$\lambda$.  Then, we obtain
from (\ref{eq:current0})
\begin{eqnarray}
(-f^{-1}f'A_0' + A_0'')\nabla{\lambda}\cdot\nabla{\lambda} +
A_0'\nabla^2\lambda=0,\label{eq:aux2}
\end{eqnarray}
where $()'$ denotes derivatives respect to $\lambda$. Whereas, from (\ref{eq:auxiliar}),
we have
\begin{eqnarray}
 -f^{-1}f'\nabla{\lambda}\cdot\nabla{\lambda} + \nabla^{2}\lambda = 0 . \label{eq:aux3}  
\end{eqnarray}
Consequently, by inserting (\ref{eq:aux3}) into (\ref{eq:aux2}), we obtain 
$
 A_0'' =0
$, whose general solution is
$
 A_0 = k_1\lambda + k_2
$, where $k_1$ and $k_2$ are constants.  We now proceed to determine the function $\lambda(r,z)$.
Let us assume in (\ref{eq:aux3}) the 
very useful simplification 
$ 
 f'f^{-1}=k
$, 
where $k$ is an arbitrary constant. Then, 
$ f=k_3e^{k \lambda},
$
and 
\begin{eqnarray}
\nabla^{2}\lambda = k \nabla{\lambda}\cdot\nabla{\lambda}
,\label{eq:lambdalaplace}
\end{eqnarray}
where $k_3$ is  a constant. 
Furthermore, if we now assume the existence of  a
function $U= k_4e^{-k\lambda} + k_5$, with $k_4$ and $k_5$ being arbitrary constants, then 
\begin{eqnarray}
 \nabla^2{U} = -kk_4e^{-k\lambda}(\nabla^2\lambda
-k\nabla\lambda\cdot\nabla\lambda) = 0
\end{eqnarray}
and, consequently, $\lambda$ can be represented in terms of  solutions of the
Laplace  equation:
\begin{eqnarray}
 e^{k\lambda}=\frac{k_4}{U-k_5};\qquad \nabla^2U=0 \label{eq:metlap}.
\end{eqnarray}
Then, in order to have an asymptotically flat
spacetime at infinity, we will only consider functions $U$ that vanish at infinity and we
must take $k_5=-k_3 k_4$ in  Eq. (\ref{eq:metlap}).

On the other hand, from Eq.(\ref{eq:gradalamb})  we obtain the following relationship between $A$ and $\lambda$:
\begin{eqnarray}
&& \nabla{A}=A_{,r}\hat{\mathbf e}_{r} + A_{,z}\hat{\mathbf e}_{z}=
 rf^{-1}\hat{\mathbf e}_{\varphi}\times(\lambda_{,r}\hat{\mathbf e}_{r} +
\lambda_{,z}\hat{\mathbf
e}_{z}),
\end{eqnarray}
that is, 
$
A_{,r}=-rf^{-1}\lambda_{,z},$ and 
$A_{,z}=rf^{-1}\lambda_{,r}$,
or, in terms of $U$,
$
A_{,r}=k_6rU_{,z}$, and
$A_{,z}=-k_6rU_{,r}$,
where $k_6= 1/(kk_3k_4)$. Then, using this solution for the
nonzero components of ${M_{(a)(b)}^{\pm}}$, we have from (\ref{eq:emth2m}):  
\begin{subequations}\begin{eqnarray}
{M_{(0)(0)}^{\pm}} &=&
\frac{U_{,r}^2+U_{,z}^2}{4(U - k_5)^2}\{3f - k_7\},\\
{M_{(1)(1)}^{\pm}} &=&
\frac{U_{,r}^2+U_{,z}^2}{4(U - k_5)^2}\{f - k_7\},\\
{M_{(2)(2)}^{\pm}} &=&
-\frac{U_{,r}^2-U_{,z}^2}{4(U - k_5)^2}\{f - k_7\},\\
{M_{(2)(3)}^{\pm}} &=&
-\frac{U_{,r}U_{,z}}{2(U - k_5)^2}\{f - k_7\},\\
{M_{(3)(3)}^{\pm}}&=&
-{M_{(2)(2)}^{\pm}}.
\end{eqnarray}\end{subequations}
Furthermore, for the nonzero components of the energy-momentum tensor and the current density on the
surface of the  disk we have from (\ref{eq:stet}) and (\ref{eq:currtet})
respectively,
\begin{eqnarray}
 S_{(0)(0)}= 2(k_3k_4)^{1/2}\frac{U_{,z}}{(U-k_5)^{3/2}}\label{eq:s00}
\end{eqnarray}
and
\begin{subequations}
 \begin{eqnarray}
  {\cal J}_{(0)}&=&\frac{k_1}{k}\left[\frac{U_{,z}}{U-k_5}\right],\\
  {\cal J}_{(1)}&=&\frac{[U_{,r}]}{k(U-k_5)},\label{eq:j0j1}
 \end{eqnarray}
\end{subequations} 
with
\begin{eqnarray}
f= \frac{k_3k_4}{U - k_5}\qquad \mbox{and}\qquad
k_7 =  \frac{2(k_1^2 + 1)}{k^2}.\end{eqnarray}
As we  can
see, $S_{(0)(0)}$ is  the only nonzero component of the surface energy-momentum 
tensor; we can then interpret it as the surface energy density of the disk
$\epsilon(r)$ as seen from any LSO. So, in order that the surface energy-momentum tensor will be in agreement with the energy conditions, we only need to require that $\left. U_{,z} \right|_{0^+} \geq 0$ and take $k_3 k_4 > 0$. Likewise, we  interpret $\sigma(r)= {\cal J}_{(0)}$
and ${\cal I}= {\cal J}_{(1)}$  as  the charge density and electric current density on the
surface of the disk, respectively.
\subsection{The eigenvalue problem for the energy-momentum  tensor  of the halo}

When the tensor ${M_{(a)(b)}^{\pm}}$ in the LSO
tetrad  is diagonal, its interpretation is immediate. In our case, however, 
${M_{(2)(3)}^{\pm}}\neq0$ and, therefore, it is necessary to rewrite ${M_{(a)(b)}^{\pm}}$ in the canonical form. 
To this end, we must solve the eigenvalue problem for ${M_{(a)(b)}^{\pm}}$, i.e.,
\begin{eqnarray}
 {M_{(a)(b)}^{\pm}}\xi^{(b)}_{\;\;\;\;\;A}
&=& \lambda_A\eta_{(a)(b)}\xi^{(b)}_{\;\;\;\;\;A}\ ,
\end{eqnarray}
and express the  physically
relevant quantities in terms of the eigenvalues and eigenvectors. 
The solution of the eigenvalue problem in the LSO  orthonormal tetrad
leads to the eigenvalues
\begin{subequations}
 \begin{eqnarray}
\lambda_0 &=& - {M_{(0)(0)}^{\pm}},\\
\lambda_1 &=&  {M_{(1)(1)}^{\pm}},\\
\lambda_{\pm} &=& \pm \sqrt{D},
 \end{eqnarray}
\end{subequations}
and the  corresponding  eigenvectors  are given by
\begin{subequations}
 \begin{eqnarray}
\xi^{(a)}_{\;\;\;\;\;0} &=& V^{(a)}= (1,0,0,0),\\
\xi^{(a)}_{\;\;\;\;\;1} &=& X^{(a)}=  (0,1,0,0),\\
\xi^{(a)}_{\;\;\;\;\;+} &=& Y^{(a)}=N(0,0,1,-\omega),\\
\xi^{(a)}_{\;\;\;\;\;-} &=& Z^{(a)}=N(0,0,\omega,1),
 \end{eqnarray}
\end{subequations}
where
\begin{eqnarray*}
D&=&({M_{(2)(2)}^{\pm}})^2 + ({M_{(2)(3)}^{\pm}})^2,\\
\omega&=&\frac{{M_{(2)(2)}^{\pm}} - \sqrt{D}}{{M_{(2)(3)}^{\pm}}},\\
N&=&\frac{1}{\sqrt{1 + \omega^2}}.
\end{eqnarray*}
In terms  of  the tetrad 
$
 \xi^{(a)}_{\;\;\;\;A}=\{V^{(a)}, X^{(a)}, Y^{(a)}, Z^{(a)} \},
$
the tensor  ${M_{(a)(b)}^{\pm}}$  can be  written in the canonical  form
\begin{eqnarray}
 {M_{(a)( b )}^{\pm}}=
 \varepsilon V_{( a )}V_{( b )} + p_1 X_{( a )}X_{( b )} +
 p_2Y_{( a )}Y_{( b )} + p_3 Z_{( a )}Z_{( b )}.
\end{eqnarray}
Consequently, we can interpret $\varepsilon$ as the energy density of the halo and
$p_1,p_2$ and $p_3$ as the pressure in the principal directions of the  halo. So, we 
have that the energy density of  the halo is  given by
\begin{equation}
\varepsilon= {M_{(0)(0)}^{\pm}} \label{eq:enhalo}
\end{equation}
whereas for the principal pressure  we have
\begin{equation}
 p=p_1=p_2=-p_3={M_{(1)(1)}^{\pm}},\label{eq:presshalo}
\end{equation}
and
\begin{eqnarray}
\langle p \rangle = \frac{p_1 + p_2 + p_3 }{3}= \frac p3
\end{eqnarray}
is  the average value of the pressure.

Accordingly, we have an anisotropic fluid with a non-barotropic equation of state, which can be written as
\begin{equation}
p = p (f, \varepsilon) =\left[ \frac{f - k_7}{3 f - k_7} \right] \varepsilon,
\end{equation}
in such a way that the pressure not only depends on the energy density but also on the gravitational and electromagnetic fields through the function $f$. Now, it is easy to see that, in order that the material energy-momentum tensor satisfy all the energy conditions, the function $U$ must be constrained by
\begin{equation}
U \leq \frac{(2 - k_7)k_3 k_4}{k_7},
\end{equation}
where we used the constrain on $k_5$ in  Eq. (\ref{eq:metlap}) needed to have an asymptotically flat
spacetime at infinity. 

\subsection{The Kuzmin-like solution}

One  can find  many different models for a relativistic thin disk surrounded by a
material electromagnetized halo by choosing different kind of solutions $U$ of the Laplace
equation. Let us consider the particular case of Kuzmin's solution \cite{KUZMIN, BT}
\begin{eqnarray}
 U=-\frac{m}{\sqrt{r^2 + (|z| + a)^2}},\qquad (a, m > 0).\label{eq:uk}
\end{eqnarray}
At points with $z < 0$, $U$ is identical to the
potential of a point mass $m$ located at the point $(r, z) = (0, -a)$, and when
$z > 0$, $U$ coincides with the potential generated by a point mass at $(0, a)$.
Hence $\nabla^2 U$ must vanish everywhere except on the plane $z = 0$. By applying
Gauss theorem to a flat volume that contains a small portion of the
plane $z = 0$, we conclude that $U$ is generated by the surface density of a Newtonian mass
\begin{eqnarray}
 \rho(r,z=0)=\frac{am}{2\pi(r^2 + a^2)^{3/2}}.
\end{eqnarray}
Furthermore, without loss of 
generality we can choose the constants $k_3=k_4=1$. Then, for the metric potential we have
\begin{eqnarray}
 e^{2\phi}=\frac{\sqrt{\tilde r^2 + (|\tilde z| + 1)^2}}{ {\sqrt{\tilde r^2 + (|\tilde z|
+ 1)^2}} - {\tilde m} } \ ,\label{eq:ksol}
\end{eqnarray}
where  we introduced the dimensionless variables  $ \tilde r=r/a$, $\tilde{z} = z/a$, and  $ \tilde{m}
= m/a$. With this metric  potential it is straightforward  to 
calculate the  dimensionless radial and axial components of the electric field, ${\tilde
E}_r =a E_r$ and ${\tilde E}_z =a^2 E_z$, and the dimensionless radial and axial components
of the magnetic field, ${\tilde B}_r = B_r$ and ${\tilde B}_z =B_z$. We obtain 
\begin{subequations}
 \begin{eqnarray}
 {\tilde E}_r&=& -\frac{k_1\tilde m\tilde{r}}{k\left\{\tilde{r}^2 +
(|\tilde{z}|+1)^2\right\}
\left\{\sqrt{\tilde{r}^2 + (|\tilde{z}|+1)^2} - \tilde{m}\right\}},\\
 {\tilde E}_z&=& -\frac{k_1\tilde m|\tilde{z}|(\tilde{z} +
1)}{k\tilde{z}\left\{\tilde{r}^2 
 + (|\tilde{z}|+1)^2\right\}\left\{\sqrt{\tilde{r}^2 + (|\tilde{z}|+1)^2} -
\tilde{m}\right\}},
 \\
 {\tilde B}_r&=& -\frac{{\tilde m}{\tilde r}^2}{k\left\{{\tilde r}^2 
+ (|\tilde{z}|+1)^2\right\}^{3/2}},\\
{\tilde B}_z&=& -\frac{{\tilde m}{\tilde r}{|\tilde z|}({\tilde z} +1)}
{k{\tilde z}\left\{{\tilde r}^2 + (|\tilde{z}|+1)^2\right\}^{3/2}}, 
\end{eqnarray}
\end{subequations}
were we have  used the definitions $E_r = A_{0,r}, E_z = A_{0,z}, B_r = A_{,z}$ and $B_z =
- A_{,r}.$

Substituting (\ref{eq:uk})   into (\ref{eq:s00}) and (\ref{eq:j0j1}),  we obtain for
the dimensionless energy and charge density on surface of  the  disk, $\tilde{\epsilon} =
a\epsilon$, $\tilde{\sigma} = a\sigma$, respectively,
\begin{subequations}\begin{eqnarray}
\tilde\epsilon(\tilde{r}) &=& \frac{2\tilde{m}}{(\tilde{r}^2 + 1)^{3/4}
(\sqrt{\tilde{r}^2 + 1} - \tilde{m})^{3/2}},\\ 
\tilde\sigma(\tilde{r}) &=&\frac{2k_1\tilde{m}}{k(\tilde{r}^2 + 1)
(\sqrt{\tilde{r}^2 + 1} - \tilde{m})}.
\end{eqnarray}\end{subequations}
The substitution of (\ref{eq:uk}) into (\ref{eq:enhalo}) and
(\ref{eq:presshalo}) allows us to write  the dimensionless energy of the
halo
$
 \tilde{\varepsilon}(\tilde{r},\tilde{z}) = a^2{\varepsilon}(r,z)$ and the dimensionless pressure 
of  the halo
$ \tilde{p}(\tilde{r},\tilde{z}) = a^2{p}(r,z)$ as
\begin{subequations}
\begin{eqnarray}
\tilde{\varepsilon}(\tilde{r},\tilde{z})= \tilde{T}\left\{ \frac{3\sqrt{\tilde{r}^2 +
(|\tilde{z}|+1)^2}}{\sqrt{\tilde{r}^2 + (|\tilde{z}|+1)^2} -
\tilde{m}} - k_7\right\},\\
\tilde{p}(\tilde{r},\tilde{z})= \tilde{T}\left\{ \frac{\sqrt{\tilde{r}^2 +
(|\tilde{z}|+1)^2}}{\sqrt{\tilde{r}^2 + (|\tilde{z}|+1)^2} -
\tilde{m}} - k_7\right\},
\end{eqnarray}
\end{subequations}
where
\begin{eqnarray*}
\tilde{T}=\frac{\tilde m}{4\left\{\tilde{r}^2 + (|\tilde{z}|+1)^2\right\}
\left\{\sqrt{\tilde{r}^2 + (|\tilde{z}|+1)^2} - \tilde{m}\right\}^2}.
\end{eqnarray*}

To fix the values of the constants that enter the solution, we consider the 
energy conditions at infinity, $\tilde r,\tilde z\longrightarrow \infty$, and at 
the center of the body, $\tilde r,\tilde z\longrightarrow 0$. Then, it can be  shown
that all the energy conditions  are satisfied at these extreme regions if
\begin{equation}
\tilde m < 1
\end{equation}  
for the disk and, additionally, 
\begin{equation}
-\sqrt{k_1^2 + 1} <  k < \sqrt{k_1^2 + 1} \qquad (\mbox{that is}\;\;\;
k_7<2)
\end{equation}
for the halo.


\begin{figure}
$$\begin{array}{cc}
{\tilde \epsilon} & {\tilde
\sigma}\\
\epsfig{width=3in,file=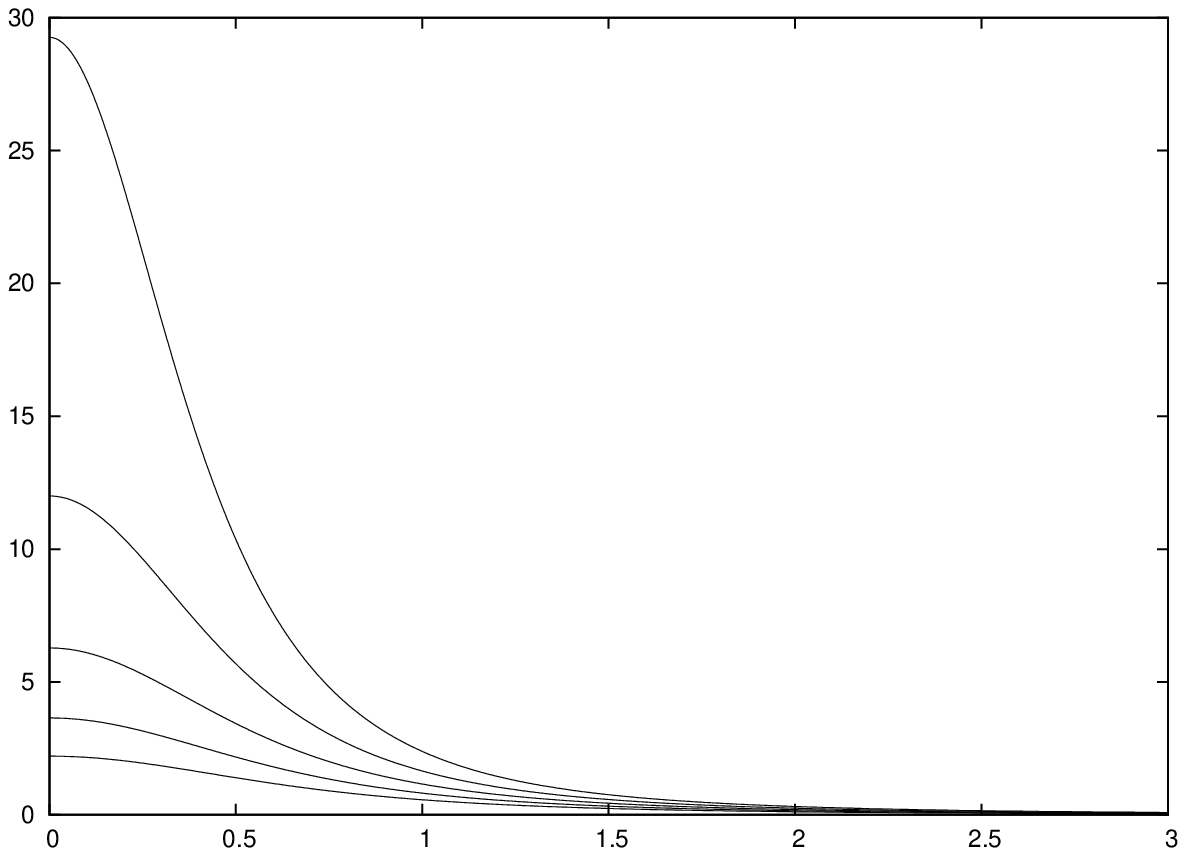}&
\epsfig{width=3in,file=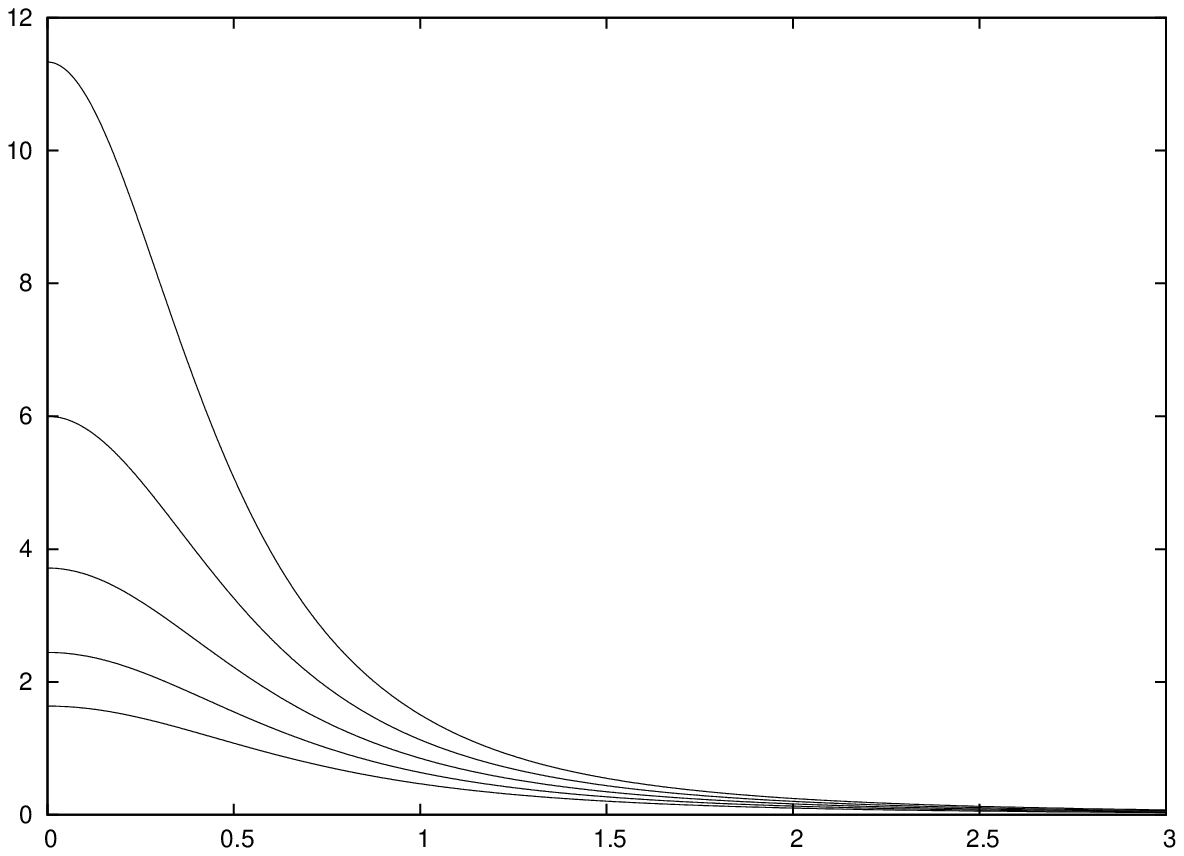}\\
\tilde{r} & \tilde{r}\\
&\\
(a) & (b)\\
\end{array}$$
\caption{\label{fig:figure1} Dimensionless surface energy ${\tilde \epsilon}$ and charge
${\tilde \sigma}$ densities as a function of ${\tilde r}$. In each case, we plot ${\tilde
\epsilon}(\tilde r)$ and ${\tilde \sigma}(\tilde r)$ for different values of the parameter
$\tilde{m}$. First, we take $ \tilde{m} =0.45$ (the bottom curve in each plot) and then
$0.55,\;0.75$ and $\tilde m = 0.85$ (the top curve in each plot).}
\end{figure}
In Fig. \ref{fig:figure1}(a), we show the dimensionless surface energy density on
the disk ${\tilde \epsilon}$  as a function of ${\tilde r}$ and for different values of the parameter $\tilde{m}$. 
First, we take $
\tilde{m} =0.45$ (the bottom curve in the plot) and then $0.55,\;0.75$ and $\tilde m =
0.85$ (the top curve in the plot). It can be seen that the energy density is everywhere
positive fulfilling the energy conditions.  It can be observed that for all the values of
${\tilde m}$ the maximum of the energy density occurs at the center of the disk 
and that
it vanishes sufficiently fast as $r$ increases. It can also be observed that the energy
density in the central region of the disk increases as the values of the parameter $\tilde
m$ increase. We have also  plotted in Fig. \ref{fig:figure1}(b) the charge  density
${\tilde \sigma}$ as a function of ${\tilde r}$. In each case, we plot ${\tilde \sigma}$
for different values of the parameter $\tilde{m}$. We observe that the electric charge
density has a behavior similar to that of the energy. This is consistent with the fact that the
mass and the charge are more densely concentrated in the center of the disk. We also computed
these functions for other values of the parameters ${\tilde m}$  in the interval $(0,1)$ and in all cases 
we found a similar behavior.

\begin{figure}
$$\begin{array}{cc}
\epsfig{width=3in,file=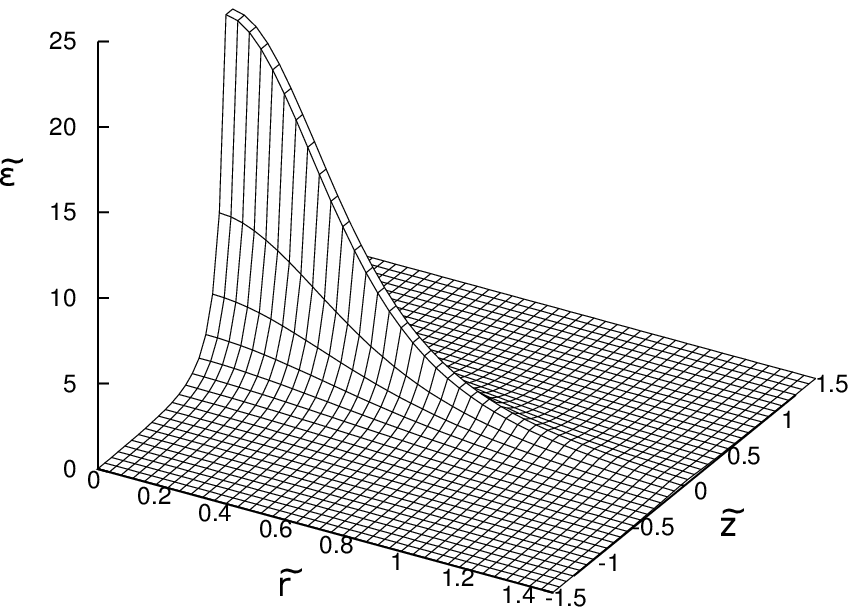}&
\epsfig{width=2.5in,file=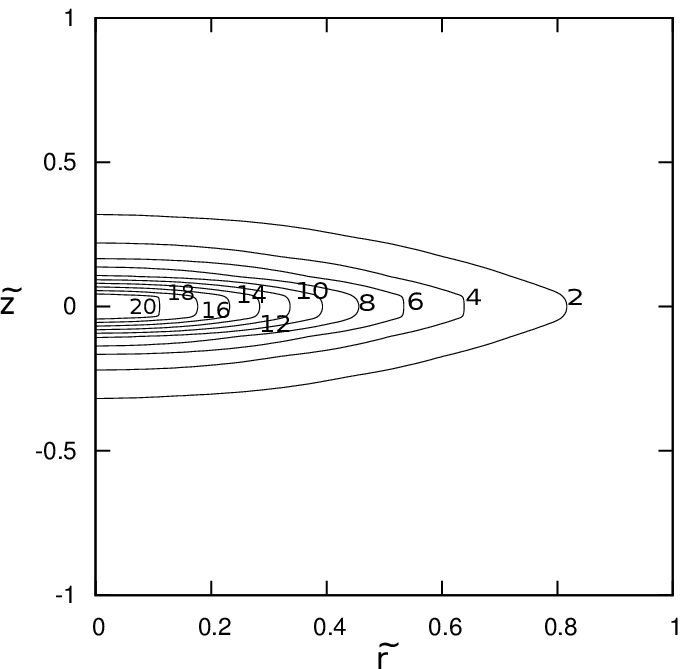}\\
&\\
(a) & (b)\\
\end{array}$$
\caption{\label{fig:figure2} Surface plot an level curves of  the energy density ${\tilde
\varepsilon}$ on the exterior halo  as a function of ${\tilde r}$ and
$\tilde z$ with parameters  $\tilde{m} = 0.75$ and $\tilde{k_7} = 1$.}
\end{figure}
\begin{figure}
$$\begin{array}{cc}
\epsfig{width=3in,file=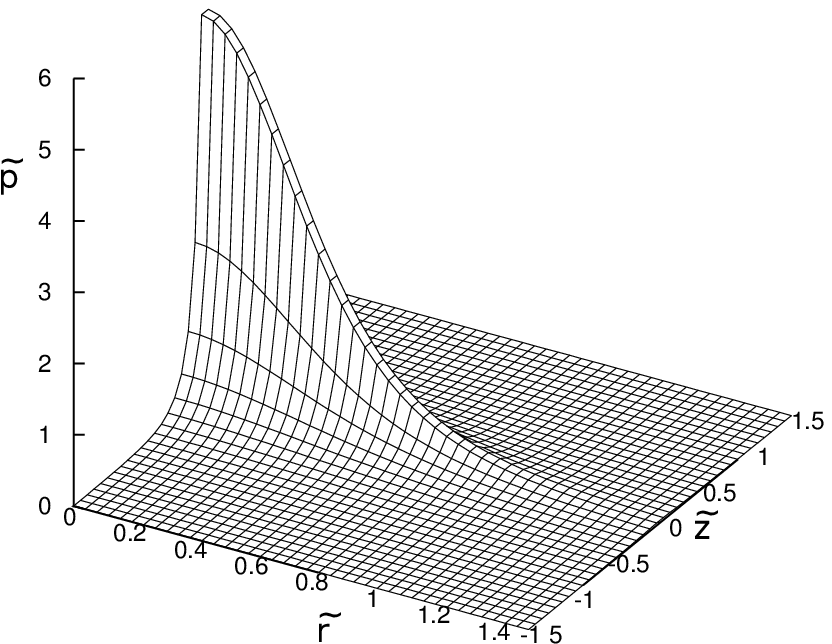}&
\epsfig{width=2.5in,file=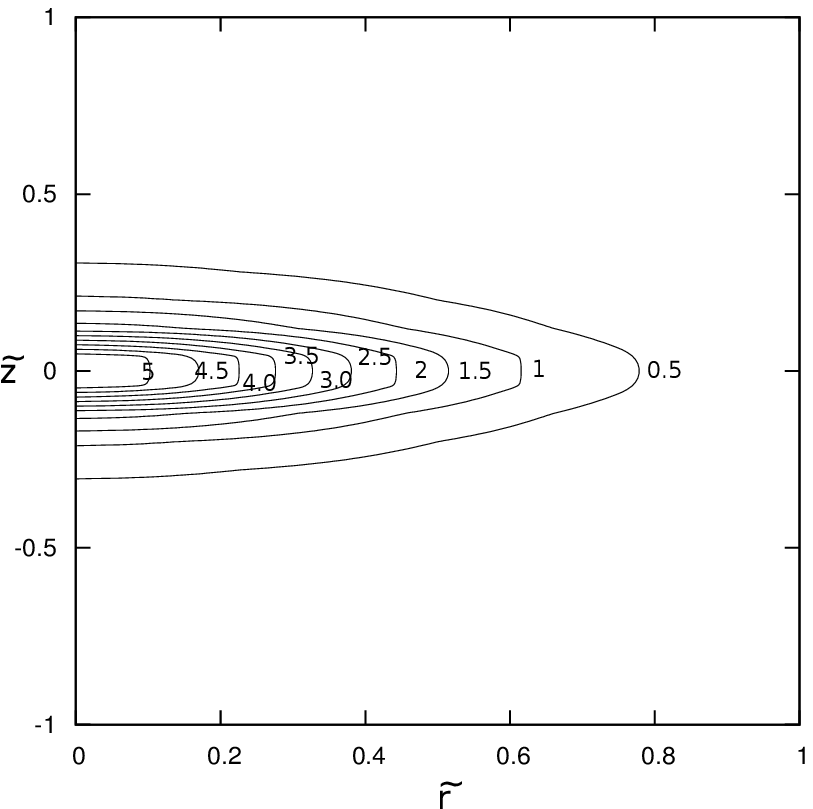}\\
&\\
(a) & (b)\\
\end{array}$$
\caption{\label{fig:figure3} Surface plot and level curves of  the  radial pressure
${\tilde p}$ on the exterior halo  as a function of ${\tilde r}$ and
$\tilde z$ with parameters  $\tilde{m} = 0.75$ and $\tilde{k_7} = 1$.}
\end{figure}
\begin{figure}
$$\begin{array}{cc}
\epsfig{width=3in,file=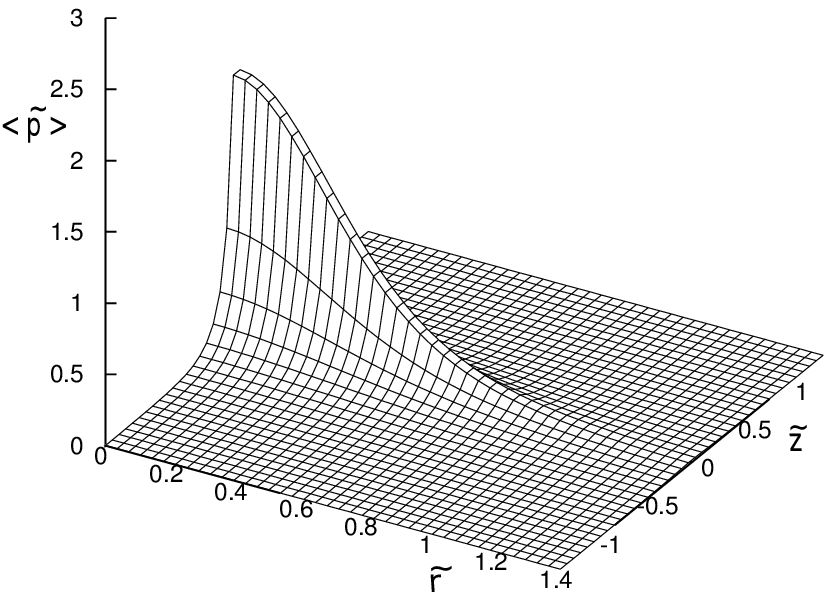}&
\epsfig{width=2.5in,file=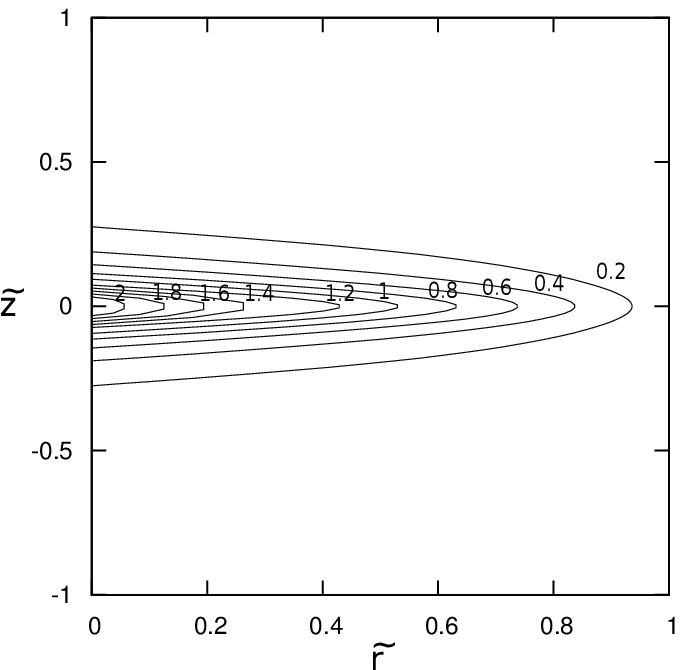}\\
&\\
(a) & (b)\\
\end{array}$$
\caption{\label{fig:figure4} Surface plot an level curves of  the average value of the radial
pressure
$<{\tilde p}>$ on the exterior halo  as a function of ${\tilde r}$ and
$\tilde z$ with parameters  $\tilde{m} = 0.75$ and $\tilde{k_7} = 1$.}
\end{figure}

In Fig. \ref{fig:figure2}(a) and in Fig.
\ref{fig:figure2}(b), we illustrate the behavior of the surfaces and level curves of the matter density in the
halo around of  the disk ($r\geq 0, z\geq 0$) for the parameters ${\tilde m} = 075$ and $k_7
=1$. We  can  see that the energy is everywhere positive, its maximum 
occurs around  the center of the disk, and it vanishes sufficiently fast as
$r$ increases.

Notice that in the limit $ m\rightarrow 0$, the gravitational and electromagnetic fields vanish identically, and the metric becomes 
flat. This is an important limiting case because it  indicates that 
$m$ determines the mass of the disk and the halo, and that the electromagnetic field exists only in connection with the disk-halo 
configuration.  Moreover, when $k_1 =0$,  we obtain  $A_0 = k_2=$ const.,  a ``purely magnetic''  solution. Then, $k_1$ determines the electric
charge of the distribution whereas $k_6 = 1 / k$ must be associated with the magnetic field. 

The energy condition $\tilde m <1$ or, equivalently, $m<a$ imposes a maximum on the value of the mass parameter. Recall that in the Kuzmin solution $a$ represents the distance along the axis between the equatorial plane and the point where the mass $m$ is situated; accordingly, $a$ can be interpreted as a parameter  determining a proper length for the configuration. Then, the inequality $m/a<1$ represents a condition on the ``specific mass" of the system. This resembles the well-known Chandrasekhar limit for a spherically symmetric mass distribution, stating that the condition mass/radius $<4/9$ must be satisfied in order to avoid gravitational collapse. The energy condition for the halo $k^2<1+k_1^2$ also represents a relationship between the parameters that characterize the electric and
magnetic properties of the system.    

We conclude that all the relevant quantities show a physically reasonable behavior within the
allowed range of values of the parameters. This indicates that the solution presented here 
can be used to describe the gravitational field of a static thin disk surrounded by a material halo 
with a non-trivial electromagnetic field.

\subsection{Singular behavior of  the  Kuzmin-like solution}
\label{sec:beh}

In order to study the singularities that could be present in the space described by the 
Kuzmin-like solution derived in the last subsection, we compute the most important quadratic curvature 
scalars, namely, the Kretschmann ${\cal K}_I$, the Chern-Pontryagin ${\cal K}_{II}$ and the Euler invariants ${\cal K}_{III}$ 
defined as \cite{cbcr02} 
\begin{subequations}
 \begin{eqnarray}
  {\cal K}_{I}&=&R^{abcd} R_{abcd},\\
  {\cal K}_{II}&=&[^*R]^{abcd} R_{abcd} = \frac{\epsilon^{ab}_{\ \ ij}R^{ijcd} R_{abcd} }{\sqrt{-g}},\\
  {\cal K}_{III}&=&[^*R^*]^{abcd} R_{abcd} = \frac{\epsilon^{abij} \epsilon^{cdkl}R_{ijkl}R_{abcd}}{g}.
 \end{eqnarray}
\end{subequations}
As for the Maxwell field, we consider the electromagnetic invariants 
\begin{subequations}
\begin{eqnarray}
{\cal F}_{_{I}}&=& F_{ab}F^{ab}\\
{\cal F}_{_{II}}&=& F_{ab}F^{*ab}.
\end{eqnarray}
\end{subequations}
Here $g=\mbox{det}(g_{ab})$,  $\epsilon^{abcd}$ is  the
Levi-Civita symbol and the asterisk denotes the dual operation.
By using the solution (\ref{eq:ksol}) we can
cast these  invariants as
\begin{subequations}
\begin{eqnarray*}
{\cal K}_{I}({\tilde r},{\tilde z})&=& 
\frac{{\tilde m}^2\left\{48\left[(|{\tilde z}| + 1)^2 + {\tilde r}^2\right]  - 32{\tilde
m}\sqrt{(|{\tilde z}| + 1)^2 + {\tilde r}^2} 
+ 11{\tilde m}^2\right\}}{4\left[(|{\tilde z}| + 1)^2 + {\tilde
r}^2\right]\left[\sqrt{(|{\tilde z}| + 1)^2 + {\tilde r}^2} - {\tilde m}\right]^6}
,\\
  & \\
{\cal K}_{II}({\tilde r},{\tilde z})&=&0,\\
& \\
{\cal K}_{III}({\tilde r},{\tilde z})&=&  
\frac{16\tilde{m}^2\left\{3[(|\tilde{z}| + 1)^2 + \tilde{r}^2] -
2\tilde{m}\sqrt{(|\tilde{z}| +
1)^2 + \tilde{r}^2}\right\}}{[(|\tilde{z}| + 1)^2 + \tilde{r}^2]\left[\sqrt{(|\tilde{z}| + 1)^2 +
\tilde{r}^2} - \tilde{m}\right]^6}
\ ,
\end{eqnarray*}
\end{subequations}
whereas the electromagnetic  invariants are
\begin{subequations}
\begin{eqnarray}
{{\cal F}}_{_{I}} =\frac{2(1 - k_1^2){\tilde
m}^2a^2}{k^2\left[(|{\tilde z}| + 1)^2 + {\tilde r}^2\right]\left[\sqrt{(|{\tilde z}| + 1)^2
+ {\tilde r}^2} - {\tilde m}\right]^2}\ ,\label{eq:elinv1}\\
 {{\cal F}}_{_{II}} =\frac{-4k_1{\tilde
m}^2a^2}{k^2\left[(|{\tilde z}| + 1)^2 + {\tilde r}^2\right]\left[\sqrt{(|{\tilde z}| + 1)^2
+ {\tilde r}^2} - {\tilde m}\right]^2}.
\end{eqnarray}
\end{subequations}

We see that 
there exists a singularity at the surface determined by the equation
\begin{equation}
{(|{\tilde z}| + 1)^2
+ {\tilde r}^2} = {\tilde m}^2 \ ,
\end{equation}
where all the non-trivial invariants diverge. The singularity equation allows real solutions only for $\tilde m > 1$. 
On the other hand, in the previous subsection we showed that 
the energy condition for the disk-halo configuration implies that $\tilde m <1$. 
In fact, we can see from the expressions for the matter and charge density
of the halo that they both diverge at the singular surface. Such divergencies are usually associated with the collapse of the gravitational configuration. 
This is an interesting result because it shows that the spacetime becomes singular as soon as the energy condition is violated. From a physical point of view this means
that a disk-halo system can be constructed only if the condition $m/a<1$ is satisfied. As noticed in the last subsection, this condition implies an upper bound on the value of the mass parameter $m$ which is determined by the position of the mass along the symmetry axis. This, again, might be interpreted as a Chandrasekhar-like limit for the disk-halo system.

\section{Concluding remarks}\label{sec:conc}
In this work, we derived a relativistic model describing a thin disk surrounded by a halo in
presence of an electromagnetic field. The model was obtained by solving the Einstein-Maxwell equations 
on a particular conformastatic spacetime in which only one independent metric function appears.   
For the energy-momentum tensor we used the distributional approach,  and represented it as 
the sum of two distributional contributions, one due to the electromagnetic part and the other associated with a matter distribution. 
These assumptions allowed us to derive explicit expressions for the energy, pressure, electric current and electromagnetic
field of the disk region and  the halo as well.  The main point of this approach is that it allows to 
write the gravitational and electromagnetic potentials in terms of a solution of Laplace's equation.

As a particular example, we used one of the simplest solutions of Laplace's equation, known as the Kuzmin solution,  which contains two independent 
parameters, namely, the mass  $m$ and the parameter $a$ that determines the proper length of the mass system. The resulting Kuzmin-like 
solution contains four independent parameters which determine the mass, proper length of the seed Kuzmin solution, electric charge and the magnetic field.
The solution is asymptotically flat in general and turns out to be free of singularities if the ratio $m/a<1$, an inequality that also guarantees the 
fulfillment of all the energy conditions. We interpret this condition as a Chandrasekhar-like limit for the disk-halo system.

Since all the relevant quantities show a physically reasonable behavior within the range $m/a<1$, we conclude that the solution presented here can be used to describe the gravitational and electromagnetic fields of a think disk surrounded by a halo in the presence of an electromagnetic field.

Furthermore, with the particular solution of Laplace's equation here choosen, the matter distribution 
both of the disk and the halo vanishes sufficiently fast as
$r$ and $z$ increase. This can be considered as indicating that the sources, the halo and
the disk, are of finite size. Now, this reasonable behavior  is a consequence of the
behavior of the Kuzmin solution, which describes a gravitational potential that have a
negative maximum at the center of the source and then decreases in absolute value,
vanishing at infinity. Accordingly, we can expect that in a more general case a similar
behavior will be obtained if solutions of the Laplace equation are considered that
correspond to concentrated Newtonian sources with decreasing gravitational potential and
gravitational field.

\section*{Acknowledgments}
One of us (A.C.G-P.) wants to thank  COLCIENCIAS, Colombia, and TWAS-Conacyt for support. GAG is supported by DIEF de Ciencias (UIS), Grant No. 5189. HQ thanks DGAPA-UNAM for support. 
This work was supported in part by DGAPA-UNAM, Grant No. 106110, and Conacyt, Grant No. 166391. 

\appendix

\section{Appendix}

For the general comformastatic metric 
\begin{eqnarray}
 \mathrm ds^2 = - \mathrm e^{2 \phi} \mathrm dt^2 \ + \mathrm e^{2 \psi}
 [r^2{\mathrm d}\varphi^2  + \mathrm dr^2 + \mathrm
 dz^2], \label{eq:met1} 
\end{eqnarray}
the nonzero components of the Einstein tensor read
\begin{subequations}\begin{eqnarray}
G_{00}^{\pm}  &=& -e^{2(\phi -\psi)}\{2(\psi_{,rr} + \psi_{,zz} + \frac{1}{r}\psi_{,r})
          + \psi_{,r}^2 + \psi_{,z}^2\},\\
G_{11}^{\pm}  &=& r^2\{ \phi_{,rr} + \phi_{,r}^2 + \phi_{,zz} + \phi_{,z}^2 + \psi_{,rr} 
+ \psi_{,zz}\},\\
G_{22}^{\pm}  &=& \phi_{,zz} + \frac{1}{r}\phi_{,r} + \phi_{,z}^2 + \psi_{,zz} 
+ \frac{1}{r}\psi_{,r} +\psi_{,r}^2 + 2 \phi_{,r}\psi_{,r},\\
G_{23}^{\pm} &=& - \phi_{,rz} - \phi_{,r} \phi_{,z} + \phi_{,r}\psi_{,z} 
+ \phi_{,z}\psi_{,r}  - \psi_{,rz} + \psi_{,r}\psi_{,z}, \\
G_{33}^{\pm} &=&  \phi_{,rr} +  \frac{1}{r}\phi_{,r} + \phi_{,r}^{2} + \psi_{,rr}
+ \frac{1}{r}\psi_{,r} + \psi_{,z}^2 + 2 \phi_{,z}\psi_{,z}\ .
\end{eqnarray}\label{eq:einstensor}\end{subequations}
Furthermore,  the nonzero components of the electromagnetic tensor (\ref{eq:tab}) are given by
\begin{subequations}\begin{eqnarray}
{E_{00}^{\pm}} &=& \frac{1}{2}e^{-2\psi}
\{A_{0,r}^2 + A_{0,z}^2 + r^{-2}e^{2(\phi - \psi)}(A_{,r}^2 + A_{,z}^2,) \},\\
{E_{01}^{\pm}} &=& e^{-2\psi}(A_{0,r}A_{,r} + A_{0,z}A_{,z} ),\\
{E_{11}^{\pm}} &=& \frac{1}{2}e^{-2\psi}
\{A_{,r}^2 + A_{,z}^2 + r^2e^{-2(\phi - \psi)}(A_{0,r}^2 + A_{0,z}^2 )\},\\
{E_{22}^{\pm}} &=&\frac{1}{2}\{-e^{-2\phi}(A_{0,r}^2 - A_{0,z}^2 ) +
r^{-2}e^{-2\psi}(A_{,r}^2 - A_{,z}^2)\},\\
{E_{23}^{\pm}} &=&-e^{-2\phi}A_{0,r}A_{0,z} + r^{-2}e^{-2\psi}A_{,r}A_{,z},\\
{E_{33}^{\pm}} &=&\frac{1}{2}\{e^{-2\phi}(A_{0,r}^2 - A_{0,z}^2)  - r^{-2}e^{-2\psi}
(A_{,r}^2 - A_{,z}^2)\}\ ,
\end{eqnarray}\label{eq:emtho}\end{subequations}
where, all the quantities are evaluated in the $z \geq 0$ and $z \leq 0$ regions.
Consequently, from the general equation  
\begin{equation}
G_{ab}^{\pm} =R^\pm_{ab} - \frac{1}{2} g_{ab} R^\pm = E^\pm_{ab} + M^\pm_{ab},
\end{equation}
and Eqs. (\ref{eq:einstensor}) and (\ref{eq:emtho}),  we
have for the nonzero components of the matter energy-momentum tensor
\begin{subequations}\begin{eqnarray}
{M_{00}^{\pm}} &=&-e^{2(\phi - \psi)} \{ 2\nabla^2\psi + \nabla\psi\cdotp\nabla\psi +
\frac{1}{2} e^{-2\phi}\nabla A_0 \cdotp\nabla A_0 + \frac{1}{2}r^{-2} e^{-2\psi}\nabla A
\cdotp\nabla A\}, \\
{M_{01}^{\pm}} &=& - e^{-2\psi} \nabla A_0 \cdotp\nabla A,\\
{M_{11}^{\pm}} &=& r^2\{ \nabla^2(\phi + \psi) - \frac{1}{r}(\phi + \psi)_{,r} + \nabla
\phi \cdotp\nabla \phi - \frac{1}{2} e^{-2\phi}\nabla A_0 \cdotp\nabla A_0 -
\frac{1}{2}r^{-2} e^{-2\psi}\nabla A \cdotp\nabla A \},\\
{M_{22}^{\pm}} &=& \nabla^2(\phi + \psi) - (\phi + \psi)_{,rr} + \phi_{,z}^2 + \psi_{r}^2
+ 2\phi_{,r}\psi_{r} + \frac{1}{2}e^{-2\phi}(A_{0,r}^2 - A_{0,z}^2) -
\frac{1}{2}r^{-2}e^{-2\psi}(A_{,r}^2 - A_{,z}^2),\\
{M_{23}^{\pm}} &=& -(\phi + \psi),_{rz} - \phi_{,r}(\phi - \psi)_{,z} + \psi_{,r}(\phi +
\psi)_{,z} + e^{-2\phi}A_{0,r}A_{0,z} - r^{-2}e^{-2\psi}A_{,r}A_{,z},\\
{M_{33}^{\pm}} &=& \nabla^2(\phi + \psi) - (\phi + \psi)_{,zz} + \phi_{,r}^2 + \psi_{,z}^2
+ 2\phi_{,z}\psi_{,z} - \frac{1}{2}e^{-2\phi}(A_{0,r}^2 - A_{0,z}^2)+
\frac{1}{2}r^{-2}e^{-2\psi}(A_{,r}^2 - A_{,z}^2).
\end{eqnarray}\label{eq:emth}\end{subequations}

Alternatively, the components of the MEMT in the local orthonormal tetrad for the general metric (\ref{eq:met1}) can be expressed as
\begin{subequations}\begin{eqnarray}
{M_{(0)(0)}^{\pm}} &=&
-e^{-2\psi}\{ 2(\psi_{,rr} + \psi_{,zz} + \frac{1}{r}\psi_{,r})
          + \psi_{,r}^2 + \psi_{,z}^2 
+ \frac{1}{2}e^{-2\phi}
(A_{0,r}^2 + A_{0,z}^2) + \frac{1}{2}r^{-2}e^{- 2\psi}(A_{,r}^2 + A_{,z}^2,) \},\\
{M_{(0)(1)}^{\pm}} &=& -r^{-1}e^{-(\phi + 3\psi)}(A_{0,r}A_{,r} 
+ A_{0,z}A_{,z} ),\\
{M_{(1)(1)}^{\pm}} &=&
e^{-2\psi}\{ \phi_{,rr} + \phi_{,r}^2 + \phi_{,zz} + \phi_{,z}^2 + \psi_{,rr} + \psi_{,zz}
-\frac{1}{2}e^{-2\phi}(A_{0,r}^2 + A_{0,z}^2 ) - \frac{1}{2}r^{-2}e^{- 2\psi}(A_{,r}^2 +
A_{,z}^2 )\},\\
{M_{(2)(2)}^{\pm}} &=& e^{-2\psi}\{\phi_{,zz} +
\frac{1}{r}\phi_{,r} + \phi_{,z}^2 + \psi_{,zz} + \frac{1}{r}\psi_{,r} +
\psi_{,r}^2 + 2 \phi_{,r}\psi_{,r} + \frac{1}{2}e^{-2\phi}(A_{0,r}^2 -
A_{0,z}^2 ) \\&-&\frac{1}{2}r^{-2}e^{-2\psi}(A_{,r}^2 - A_{,z}^2)\},\nonumber\\
{M_{(2)(3)}^{\pm}} &=& e^{-2\psi} \{- \phi_{,rz} -
\phi_{,r} \phi_{,z} + \phi_{,r}\psi_{,z} + \phi_{,z}\psi_{,r} - \psi_{,rz} +
\psi_{,r}\psi_{,z} + e^{-2\phi}A_{0,r}A_{0,z} -
r^{-2}e^{-2\psi}A_{,r}A_{,z}\},\\
{M_{(3)(3)}^{\pm}} &=& e^{-2\psi}\{\phi_{,rr} + 
\frac{1}{r}\phi_{,r} + \phi_{,r}^{2} + \psi_{,rr} + \frac{1}{r}\psi_{,r} +
\psi_{,z}^2 + 2 \phi_{,z}\psi_{,z} - \frac{1}{2}e^{-2\phi}(A_{0,r}^2 -
A_{0,z}^2)  \\&+& \frac{1}{2}r^{-2}e^{-2\psi}(A_{,r}^2 -
A_{,z}^2)\},\nonumber\\
\hat{J}^{\pm}_{(0)} &=& re^{\psi}\{ \nabla^2A_0
 - \nabla A_0\cdot\nabla(\phi-\psi)\},\\
\hat{J}^{\pm}_{(1)} &=& -e^{\phi}\{ \nabla^2A +\nabla{A}\cdot\nabla(\phi - \psi)
-\frac{2A_{,r}}{r} \}.
\end{eqnarray}\label{eq:MJtet}
\end{subequations}


\end{document}